# Learning the Representations of Atmospheric Moist Convection with Convolutional Neural Networks

Shih-Wen Tsou
Department of Atmospheric Sciences
National Taiwan University
Taipei, Taiwan
+886-2-33663941
ericakccakcc@gmail.com

Chun-Yian Su
Department of Atmospheric Sciences
National Taiwan University
Taipei, Taiwan
+886-2-33663941
b98209012@gmail.com

Chien-Ming Wu
Department of Atmospheric Sciences
National Taiwan University
Taipei, Taiwan
+886-2-33663905
mog@as.ntu.edu.tw

## ABSTRACT
The representations of atmospheric moist convection in general circulation models have been one of the most challenging tasks due to its complexity in physical processes, and the interaction between processes under different time/spatial scales. This study proposes a new method to predict the effects of moist convection on the environment using convolutional neural networks. With the help of considering the gradient of physical fields between adjacent grids in the grey zone resolution, the effects of moist convection predicted by the convolutional neural networks are more realistic compared to the effects predicted by other machine learning models. The result also suggests that the method proposed in this study has the potential to replace the conventional cumulus parameterization in the general circulation models.

## 1. INTRODUCTION
In the research of climate system on earth, clouds and convection are the most complicated and essential phenomena among all elements. The challenge remained for decades because they involved processes of multiple scales in time and space, and yet they resulted in significant energy exchange with the environment. For example, the formation of rain is caused by the collision of cloud droplets with typical diameters around 10 micrometers. On the other hand, the summertime afternoon thunderstorms can be as large as few hundred kilometers in radius. Furthermore, processes in different scales interact with each other and largely increase the complexity. The review article written by Arakawa [1] discussed the processes and uncertainties involved in great details.

### 1.1    Cloud Parameterization
General Circulation Models (GCMs) are the numerical approach to study the evolution of the earth's atmosphere through solving the Navier-Stokes equations on the sphere. The horizontal grid size of a conventional GCM is around hundreds of kilometers in their early years, and the cloud and convection processes are not able to be resolved in such a coarse scale. Therefore, subgrid-scale parameterization techniques, called cumulus parameterization [2], are developed to represent the effect of clouds and convection in finer scales. This is a very challenging problem because the grid-scale variables provide only the equilibrium states without the detailed evolution of the convective processes, and yet the subgrid phenomena post great impact on the grid-scale dynamics.

As computer power increases, the grid size used by GCMs become finer, which can better represent local to regional scale processes such as land-sea breeze over complex coastlines, and mesoscale convective systems. The finer spatial resolution can increase the variabilities of simulated convection and hence improve the performance of the modeled weather events [3]. However, with the grid size of around 10 kilometers, the advantages of utilizing high resolution are hindered by the "grey zone problem" [4]. At this spatial resolution, whether the moist convection should be parameterized or not becomes questionable. Because the clouds and convection are partially resolved; hence it becomes difficult to represent the organized convection for the lack of scale separation between convective and large-scale processes.

While GCMs aim to simulate the atmosphere of the entire earth at once, there is another type of high-resolution models in which the deep moist convection is explicitly simulated, called cloud-resolving models (CRMs). The development of cumulus parameterization in GCMs usually relies on the results of CRMs.[5][6] The statistics of the CRM output can be used to derive simple formulation to represent convective processes in the GCMs. The most common statistics used for cloud parameterization is the vertical eddy transport of moist enthalpy, $\overline{w'h'}$. This term can be interpreted as "while a warm and humid environment is accompanied by an upward motion, a convective cloud is very likely to form in the given location". The horizontal distribution of the covariance represents the location where the convection is strong. The vertical structure of moist enthalpy represents the energy input to the grid-scale.  The two features trigger further radiative and dynamic process in the atmosphere and hence are important in formulating cumulus parameterization.

Figure 1 shows the different sizes of model resolution and the phenomena they can resolve. For GCMs with grid-size around a hundred kilometers, the spatial distribution of $\overline{w'h'}$ is largely determined by the energy distribution at this grid scale because only limited phenomena can be resolved with the size of the grid. In this case, most simple statistical models can map the physical properties in GCMs to a $\overline{w'h'}$ vertical profile reasonably well. However, when the grid size goes down to the grey scale, we need better schemes to derive proper statistics which can represent more complicated phenomena.

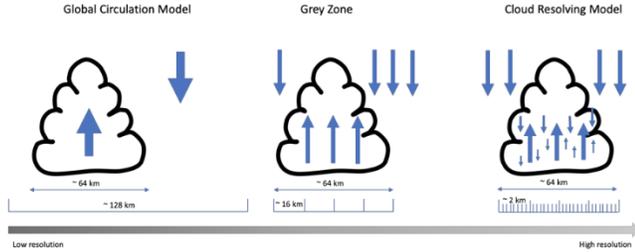

**Figure 1. An illustration of different sizes of model resolution and the phenomena they can resolve.**

## 1.2 Machine learning for cloud parameterization

Machine learning techniques are known to be able to perform a sophisticated statistical mapping between datasets. Among various learning algorithms, convolutional neural networks (CNN[5]) have been frequently applied to image recognition tasks for its being able to capture geometrical features. This characteristic makes it suitable for the datasets generated from atmospheric models.

In the studies of fluid dynamics, many advanced quantities have been proposed to merely the differential equations in the numerical models and to give a diagnosis of the equilibrium states. For example, the pressure gradient on one grid point may cause motion. These quantities usually involve the relative values of a few neighbor grids, and hence can be mathematically represented as convolutional kernels. Therefore, learning algorithms with convolutional kernels are supposed to be able to extract more informative features from the output of numerical models.

As discussed in the previous section, the task of cumulus parameterizations is to take the grid data as input variables and create a vertical profile of $\overline{w'h'}$ as the output. Hence, the neural network models become a reasonable choice for their flexibility in the output format. By using convolutional neural networks, we can output the whole vertical profile with one single model.

In this study, we try to design a cumulus parameterization scheme for GCMs with the grey zone scales with machine learning techniques. GCMs with grey zone scales can partially resolve the cloud, and convective processes and hence can provide more information to cumulus parameterization. Details of the datasets and the experiment design were described in section 2. Section 3 summarized the results and evaluation of our experiment and the significance of the proposed approach were discussed in section 4.

## 2. Methods

To investigate the feasibility of applying machine learning methods on retrieving sub-grid characteristics, a dataset similar to contemporary GCM paired with its high-resolution "ground truth" is needed. The paired datasets serve as the input and output data for machine learning models with different designs and configurations. The details of the datasets, experimental designs, and machine learning models are discussed in the following sections.

## 2.1 Datasets

To test how well machine learning models can learn the sub-grid characteristics of convective systems, we prepared two datasets with fine and coarse resolution, respectively. We used the output of a high-resolution CRM (2km)[6] as the 'truth' of the development of convective systems. The grey zone resolution (16km) dataset was generated by taking averages on 8x8 grids.

Also, in order to avoid overfitting, we want to ensure the variety of convection is covered in our training dataset. Over tropical regions, the quasi-equilibrium assumptions between convective systems and the large-scale environment can be used to generate convection dataset by providing large-scale forcing (Yanai, 1973[7]). In this approach, the statistics of convection can be obtained without producing a large number of simulations. Tsai and Wu (2017[8]) follow this idea and produce experiments by adjusting the large-scale moistening/drying to the CRM so that the environment column relative humidity ranges from 60% to 90% representing most of the possible environment in the tropics. In addition, they produce an additional set of the experiment by imposing environmental vertical wind shear so that the variety of convective organization can be captured in the dataset. In this study, we employ the whole dataset by Tsai and Wu for model training.

## 2.2 The input and output variables for machine learning

Besides the horizontal resolutions, both datasets consist of the same collection of atmospheric parameters, e.g., air pressure, geopotential height, air temperature, relative humidity, horizontal wind, ...etc. However, the complete output of the numerical model employed is too large to be analyzed point by point for each variable, and hence we chose a set of variables relevant to moist convection to evaluate.

Follow the discussion in the introduction, we use the vertical eddy transport of moist enthalpy, $\overline{w'h'}$, as the target variable. Through defining the grid-size (16 km in this study), the eddy transport of moist enthalpy can be calculated by the covariance of vertical velocity and moist enthalpy in the CRM grid size (2km in this study).

As for the input variables, we used night common variables used in atmospheric models and the study of moist convection. Six common physical variables are the horizontal wind velocity (u and v), the vertical wind velocity (w), the water vapor mixing ratio (qv), the temperature (T), and the geopotential height (z). And three thermodynamic variables are convective available potential energy (CAPE), moisture flux divergence (MFD), and moist static energy (MSE). The thermodynamic variables were known predictors for deep convection in the field of atmospheric sciences, and the values were derived from the model output. For those who are interested in the derivation of the thermodynamics variables used in this study, Petty provides an excellent introduction on the field.[11]

We used the output data described above paired with a few different subsets of the input data to verify the informativeness of the input data. The detailed experimental design is described as follows.

## 2.3 The experimental design

The primary purpose of this study is to investigate the feasibility and the proper configurations to apply machine learning on cumulus parameterization. Hence, we tested different subsets of the input data against a few different machine learning models. Table 1 listed the complete experimental design.

As shown in table 1, two sets of input data were tested: with and without the three thermodynamic variables. The purpose of this setting is to verify whether these properties from earlier research can provide more information when using a machine learning approach.

Also, we evaluate the performance of three different machine learning models, i.e., linear regression, neural networks, and convolutional neural networks. It is already well established that the dynamics of the atmosphere is highly nonlinear, and the observation of any spatial point is constrained and balanced by its neighborhood environment through physical laws. Therefore, we expect to see a performance gain by using nonlinear models and convolutional techniques.

**Table 1. the id, input variables, and model for each experiment.**

| Experiment ID | Input Variables |
|---|---|
| Linear6f | u, v, w, T, qv, z |
| Linear9f | u, v, w, T, qv, z, CAPE, MFD, MSE |
| DNN6f | u, v, w, T, qv, z |
| DNN9f | u, v, w, T, qv, z, CAPE, MFD, MSE |
| CNN6f | u, v, w, T, qv, z |
| CNN9f | u, v, w, T, qv, z, CAPE, MFD, MSE |

## 2.4 Convolutional Neural Network (CNN)

The CNN structure used in this study can be referred to as Shuiwan (2013) [12]. Figure 2 represents the model structure, which consists of two blocks. Each block contains three convolutional layers, which has 32, 64 and 128 kernels, respectively. The activation function used in this model is ReLU, followed by a max pooling layer. The output, which represents the dense layer, is then be flattened to a one-dimensional vector which has a length of five.

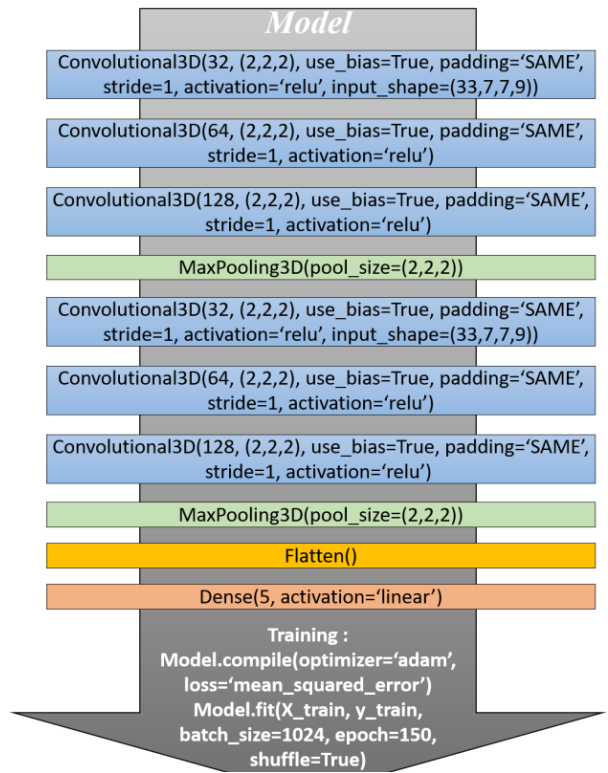

**Figure 2. The 3D-Convolutional Neural Network structure in this study.**

## 3. Results

Figure 3 shows the normalized RMSE and the correlation coefficient of each experiment. As shown in figure 3, using the nonlinear model (neural networks) outperforms the traditional linear regression, and adding convolutional kernels can further improve the learning results. Also, adding thermodynamic variables derived from the earlier study does improve the performance of every statistical model, though the improvement is relatively smaller than using superior models. In general, the experiment results meet our expectation.

Besides the general metrics for evaluating machine learning models, we also want to understand how well CNNs can learn the spatial distribution of $\overline{w'h'}$. In the next section, we will look into further details in the vertical and horizontal patterns.

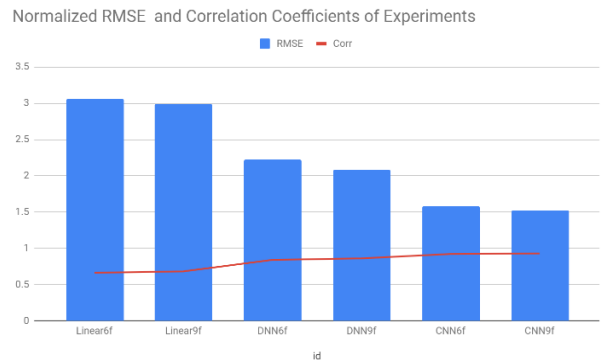

**Figure 3. The normalized RMSE and the correlation coefficient of each experiment.**

## 4. Discussion

In this section, we examine a few cases in the dataset to see what kind of spatial features the proposed method can learn.

Figure 4 shows two cases when the lower (left panel) and middle (right panel) layers are wet. As shown in the figure, the ground truth (black lines) shows kinks around 5000 meters of altitude. These characteristics represent a layer of conversion from liquid water into ice, which is common in real-world observation but usually cannot be resolved in GCMs. We can see the proposed method (3D-CNN, red lines) successfully learns this pattern while other methods cannot.

Figure 5 shows the averaged $\overline{w'h'}$ for all cases at 3000-meter altitude. The ground truth shown in the upper left panel is shown as contours in other panels. We can see the proposed approach can accurately capture the locations of convection and clouds in the horizontal profile.

In this study, we try to design a cumulus parameterization scheme for GCMs with the grey zone scales with machine learning techniques. We propose using convolutional neural networks based on the known characteristics of fluid dynamics and compare against linear models as well as nonlinear models without convolutional kernels. The results show the proposed method outperform not only other methods but also be able to capture spatial features that other methods cannot. In conclusion, we recommend using CNN and output of CRMs for cumulus parameterization.

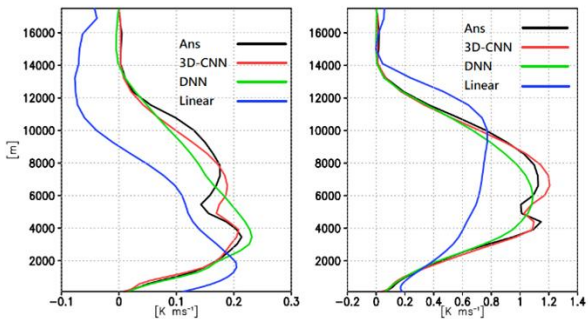

**Figure 4. The vertical profile of two cases. The case shown in the left panel has a wet lower layer, and the right panel case has a wet middle layer.**

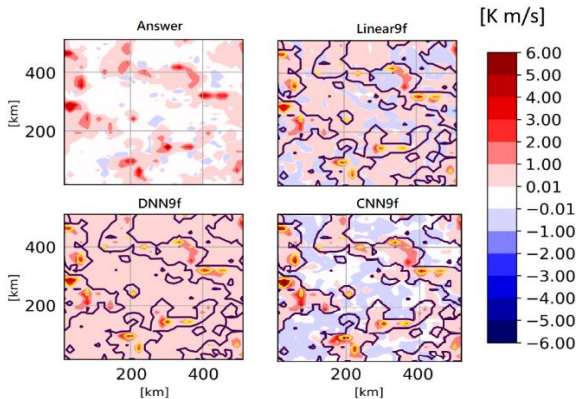

**Figure 5. The averaged for all case at 3000-meter altitude. The ground truth shown in the upper left panel is shown as contours in other panels.**

## 5. Acknowledgement

We thank Mr. Ting-Shuo Yo for in-depth discussion and useful comments on this manuscript.